\documentclass[aps,prl,preprint,superscriptaddress]{revtex4-1}

\usepackage{graphicx} 
\DeclareGraphicsExtensions{.pdf,.png,.jpg}
\usepackage{multirow}

\usepackage{amsmath}
\setcitestyle{super}

\usepackage[normalem]{ulem} 
\usepackage{color}

\begin{document}


\title{Thermally induced magnetic relaxation in square artificial spin ice}


\author{M. S. Andersson}
\email{Mikael.Andersson@angstrom.uu.se}
\affiliation{Department of Engineering Sciences, Uppsala University, Box 534, SE-751 21 Uppsala, Sweden}

\author{S.D. Pappas}
\affiliation{Department of Physics and Astronomy, Uppsala University, Box 516, SE-751 20 Uppsala, Sweden}

\author{H. Stopfel}
\affiliation{Department of Physics and Astronomy, Uppsala University, Box 516, SE-751 20 Uppsala, Sweden}

\author{E. \"{O}stman}
\affiliation{Department of Physics and Astronomy, Uppsala University, Box 516, SE-751 20 Uppsala, Sweden}

\author{A. Stein}
\affiliation{Center for Functional Nanomaterials, Brookhaven National Laboratory, P.O. Box 5000, Upton, New York 11973, USA}

\author{P. Nordblad}
\affiliation{Department of Engineering Sciences, Uppsala University, Box 534, SE-751 21 Uppsala, Sweden}

\author{R. Mathieu}
\affiliation{Department of Engineering Sciences, Uppsala University, Box 534, SE-751 21 Uppsala, Sweden}

\author{B. Hj\"{o}rvarsson}
\affiliation{Department of Physics and Astronomy, Uppsala University, Box 516, SE-751 20 Uppsala, Sweden}

\author{V. Kapaklis}
\affiliation{Department of Physics and Astronomy, Uppsala University, Box 516, SE-751 20 Uppsala, Sweden}
\date{\today}
\maketitle


{\bf The properties of natural and artificial assemblies of interacting elements, ranging from Quarks to Galaxies, are at the heart of Physics. The collective response and dynamics of such assemblies are dictated by the intrinsic dynamical properties of the building blocks, the nature of their interactions and topological constraints. Here we report on the relaxation dynamics of the magnetization of artificial assemblies of mesoscopic spins. In our model nano-magnetic system - square artificial spin ice - we are able to control the geometrical arrangement and interaction strength between the magnetically interacting building blocks by means of nano-lithography\cite{Wang:2006kta}. Using time resolved magnetometry we show that the relaxation process can be described using the Kohlrausch law\cite{Kohlrausch} and that the extracted temperature dependent relaxation times of the assemblies follow the Vogel-Fulcher law\cite{Vogel, Fulcher}. The results provide insight into the relaxation dynamics of mesoscopic nano-magnetic model systems, with adjustable energy and time scales, and demonstrates that these can serve as an ideal playground for the studies of collective dynamics and relaxations.}

Ever since Rudolf Kohlrausch's\cite{Kohlrausch} early studies of the discharge of a Leyden jar, measurements of relaxation have been a common method to gain insights into the dynamics of many-body systems. The documented examples cover a wide range of physical systems, spanning from dielectric\cite{Williams:1970ki, Nature_dielectric_relaxation_1977}, magnetic\cite{Anderson_PRL_1984}, and amorphous materials\cite{Bohmer_1993, Schoenholz:2016fb} to particles\cite{Weeks:2000dw,Andersson2014EPL} and proteins\cite{PARAK1991362}. To this day, the relaxation and diffusion in interacting many-body systems represents intriguing, yet unsolved, challenges within condensed matter physics and statistical mechanics\cite{Ngai}. 
Disorder, heterogeneity, boundaries, interfaces and presence of multiple relaxation times represent typical obstacles for a meaningful description of the relaxation in real systems.

Recently, the studies of relaxation in interacting many body ensembles have been extended to systems with inherent frustration,  caused by geometrical constraints,  such as in the spin ice pyrochlores\cite{Bramwell2001Science}. The magnetic relaxation in these magnetically frustrated materials is linked to the creation and flow of magnetic monopole currents\cite{Giblin_NatPhys_2011}, which have been found to be strongly affected by impurities and boundary effects\cite{Revell_NatPhys_2012}. In order to circumvent the obstacle of such additional complexity, we have studied the relaxation of lithographically defined nano-magnetic systems, square artificial spin ice, i.e. a two-dimensional analogue of pyrochlore spin ice materials\cite{Wang:2006kta, Heyderman_2013_review, Farhan_PRL_2013}. It was recently shown that the geometry of the lithographically defined lattices dictates their thermal dynamics\cite{Kapaklis2014}. Exploiting this result, we extend our investigation to the evolution of the magnetic relaxation of structures with different interaction energies. By determining the time dependence of the magnetization at different temperatures,  we establish a link between the micro-magnetic nature of the magnetic elements and their collective dynamic response.

\begin{figure}[h!tbp]
		\includegraphics[width=0.45\textwidth]{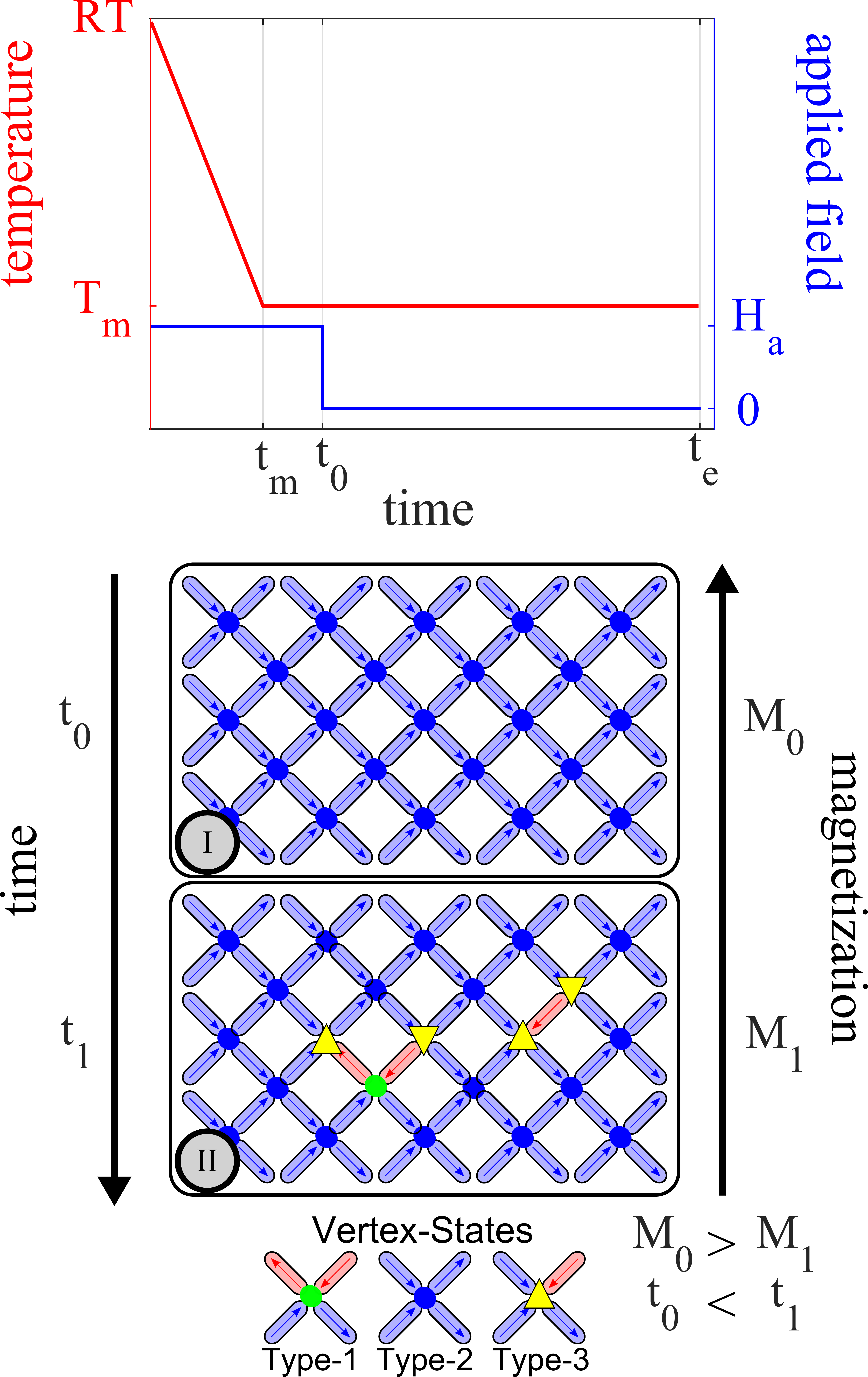}
			\centering
\caption{\textbf{Relaxation measurement protocol}. The upper graph represents the temperature and applied field evolution during the measurement protocol. Starting at room temperature (RT), where the array is in a superparamagnetic state, the external field, $H_a$, is switched on in the [110]-direction (see Fig. \ref{fig:AFM_MvsTMPMS}a) of the array and the cooling starts. As the array is cooled the magnetization of the islands aligns with the applied field.
The flipping rate of the islands is suppressed by the applied field and the low temperature. The cooling is halted at the desired measurement temperature ($T_{m}$) and stabilized there from $t_{m}$ to $t_{0}$. At $t_{0}$ the applied magnetic field is switched off and the relaxation measurements start from a completely dressed array configuration of Type-2 states (I). After the field is switched off the islands can undergo reversals and thereby reduce the magnetization of the array (II). Following the initial reversals the magnetic configuration of the array further relaxes forming strings\cite{Farhan_PRL_2013}. Three out of sixteen possible vertex states (Type-1, Type-2 and Type-3)\cite{Kapaklis2014} are shown in the bottom panel.}
	\label{fig:Intro}
\end{figure}

\begin{figure*}[htbp]
		\includegraphics[width=0.90\textwidth]{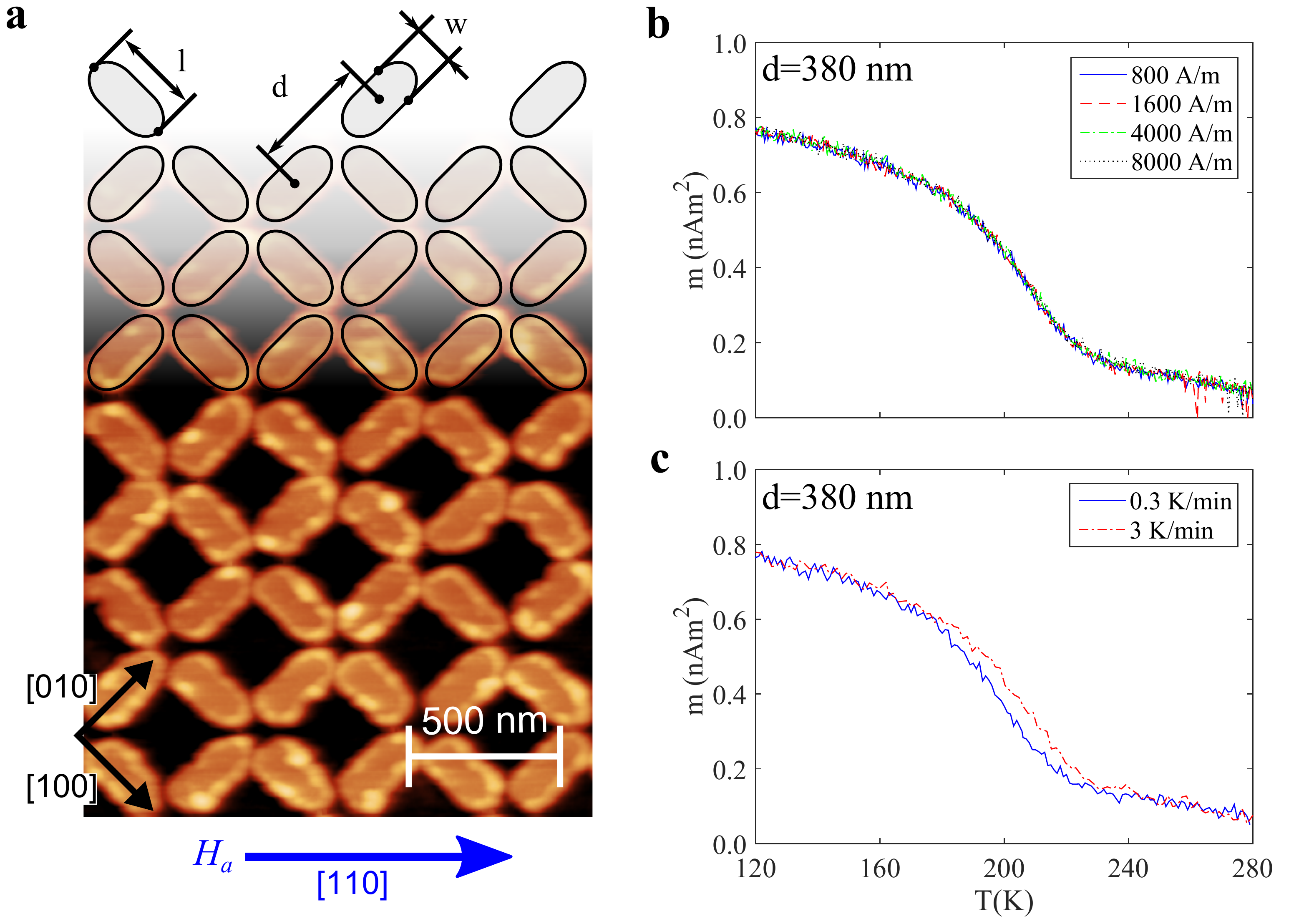}
			\centering
	\caption{\textbf{Array structure and magnetic characterization}. \textbf{a}, Atomic force microscopy image of the $d$ = 380~\rm{nm} array with an overlay showing the patterned geometry of the array. The elongated islands are stadium shaped with $l$ = $330~\rm {nm}$ and $w$ = $150~\rm{nm}$.	The islands are placed in a square lattice architecture with periodicities of $380$ $\rm{nm}$ (shown) and $420$ $\rm{nm}$ (not shown). The different perodicities lead to difference in the magnetic interaction between the islands in the arrays, with the $d$ = $380$ $\rm{nm}$ array being stronger interacting. (\textbf{b}) and (\textbf{c}) show the magnetic response of the $d$ = 380 nm array as a function of temperature. \textbf{b}, $M_{TRM}(T)$ measured after cooling in fields of different strength. As can be seen there is hardly any difference between the different curves indicating that the array starts from a fully dressed state [see Fig. \ref{fig:Intro} (I)] already at the lowest field, 800 A/m. \textbf{c}, The dependence of the $M_{TRM}(T)$ on the heating rate. The onset of decay of the collective array magnetization is shifted to higher temperatures when using a higher heating rate (i.e. a shorter observation time).  
	%
	}
	\label{fig:AFM_MvsTMPMS}
\end{figure*}

The relaxation of the square artificial spin ice was recorded using a custom built SQUID-magnetometer. By recording the time dependent magnetization, the thermal excitations of the arrays could be followed in time.  A schematic representation of the measurement protocol, the magnetic configuration of the array and plausible excitations are found in Fig. \ref{fig:Intro}. Starting from a fully dressed state, (I), all initial excitations of type-3 result in a reduction of the magnetization of the array, see Fig. \ref{fig:Intro} (II). At a next step in the relaxation process, extended strings are formed, consisting of type-3 vertices at the ends, which are connected by type-1 vertices, as illustrated in Fig. \ref{fig:Intro} (II). As for the initial excitations all of the excitations involved in the formation of a string reduce the magnetization of the array.
In this study, two arrays with different periodicity, $d$, but with the same size and geometry of the elements, were used, see Fig. \ref{fig:AFM_MvsTMPMS}a. The periodicity of the arrays is 380 nm and 420 nm, respectively. This implies a difference in interaction strength, as the distance between the elements is different in these samples. The magnetic interaction of the elements is stronger in the $d$ = 380 nm array, as compared to the $d$ = 420 nm array. This difference influences the magnetization as a function of temperature, with the transition from a frozen to a dynamic state occurring at higher temperatures for the sample with shorter distance between the elements\cite{Kapaklis2014} due to the stronger inter-island interactions. 

\begin{figure*}[htbp]
		\includegraphics[width=0.95\textwidth]{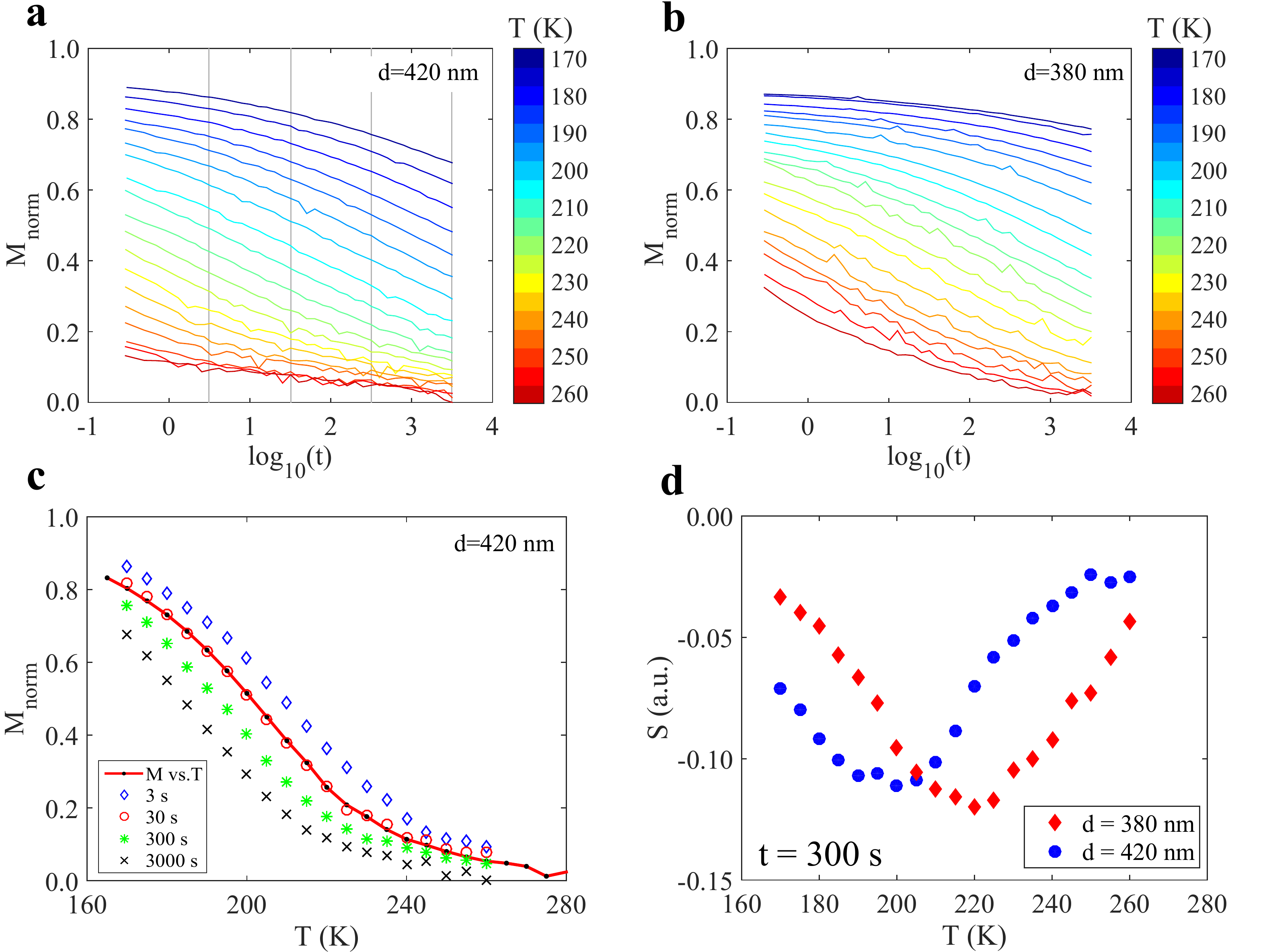}
			\centering
	\caption{\textbf{Magnetic relaxation}. Normalized magnetization as a function of time for different temperatures as indicated by the color bar for \textbf{a}, the $d$ = 420 nm array and \textbf{b}, the $d$ = 380 nm array. \textbf{c}, Normalized magnetization as a function of temperature for different observations times and the $d$ = 420 nm array. The points are selected by choosing data points along the vertical grey lines in (\textbf{a}), which correspond to specific observation times. The solid line in (\textbf{c}) corresponds to a DC M~vs.~T measurement of the $d$ = 420 nm array, the observation time of this DC measurement is 30~s. \textbf{d}, The relaxation rate, $S=dM/dlog_{10}(t)$, determined from (\textbf{a}) and (\textbf{b}), at 300~s as a function of temperature. The minimum corresponds to the maximum relaxation rate and could be taken as an effective blocking temperature for t=300 s.
	}
	\label{fig:Relax}
\end{figure*}

In magnetic relaxation experiments it is important to start from a well defined state. To ensure that the arrays are fully dressed, several magnetization measurements were performed at different temperatures, $M_{TRM}(T)$,  using different cooling fields. It was found that a field of 800 A/m (10 Oe) is sufficient to dress the array in a configuration consisting of only Type-2 states, see Fig. \ref{fig:AFM_MvsTMPMS}b. Fig. \ref{fig:AFM_MvsTMPMS}c shows two $M_{TRM}(T)$ curves for the 380 nm array measured at two different heating rates, 0.3 and 3 K/min. Increased heating rate is found to shift the transition to a higher temperature, as seen in the figure. Looking at the shape of the decay of the $M_{TRM}(T)$ curves, two features are noteworthy:  A time independent decrease in magnetization below 150 K and a time dependent decrease in the temperature range 160-230 K. The decrease in the magnetization up to 150 K is dominated by the change in the magnetization of the elements, while the contribution from excitations (reversal) of the elements is negligible at this time scale. Above 160 K, the excitations of the arrays come into play, giving rise to rate dependent reduction in the magnetization. 
 Since the thermal energy causes the excitations, the time needed for complete relaxation of an array is very long at low temperatures. On the other hand, the relaxation time will be much shorter at elevated temperatures. This in turn implies a restriction with respect to the available time window, and only a part of the relaxation of the magnetization can be obtained.
 To study the relaxation process in more detail, several temperatures in the interval 170 to 260 K were examined using the relaxation protocol described in Fig. \ref{fig:Intro} (see  also Methods). The data for the $d$ = 420 nm array is presented in Fig \ref{fig:Relax}a, where it can be seen that for low temperatures the array is still close to the dressed (saturated) state due to the slow relaxation, while at elevated temperatures the array has almost completely lost the magnetization. A similar behavior is seen for the $d$ = 380 nm array in Fig. \ref{fig:Relax}b. However, the transition for this array is shifted to higher temperatures, as seen from the relaxation curves corresponding to the $d$ = 420 nm array, which has significantly lower magnetization. It is noteworthy that the transitions for both arrays, take place at much lower temperatures than the intrinsic ordering temperature of the magnetic material, which is 410 K\cite{Kapaklis2014} (see Methods).

By selecting a specific time in the relaxation curves, the magnetization as a function of temperature for that observation time can be reconstructed. Such curves are shown for the $d$ = 420 nm array in Fig. \ref{fig:Relax}c for t=3, 30, 300 and 3000 s. 
For comparison a DC magnetization as a function of temperature measurement (M vs. T) at a heating rate of 3 K/min is also shown. As can be seen in the figure the points extracted from the relaxation measurement for 30 s and the direct DC measurement overlap.

In-between the high temperature (dynamic islands) and the low temperature (frozen islands) regions, there is an intermediate transition region where the relaxation rate is high. This is more clearly seen in Fig. \ref{fig:Relax}d which depicts the relaxation rate, $S=dM/dlog_{10}(t)$, at $t$ = 300 s, for all measured temperatures for both the arrays. The minimum in the $S(T)$ curve, corresponding to the maximum in relaxation rate, occurs at about 20 K lower in temperature for the $d$ = 420 nm array than for the 380 nm array. The temperature of the minimum for a specific time, $t$, can be used as a determination of the effective blocking temperature for the system.

\begin{figure}[t!bp]
		\includegraphics[width=0.45\textwidth]{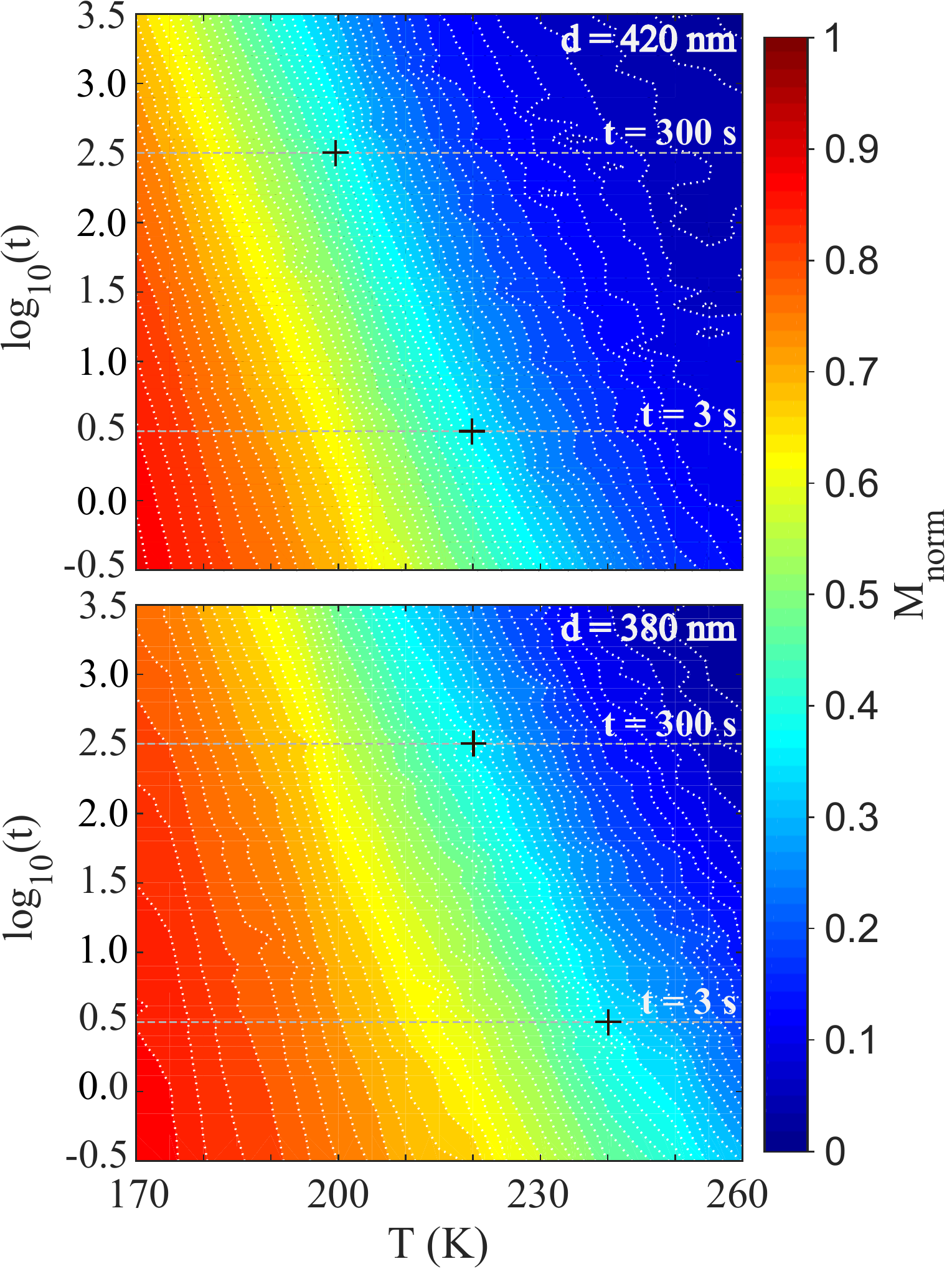}
			\centering
	\caption{\textbf{Temperature and time dependence of the magnetic relaxation}. Color maps of the magnetization as a function of time and temperature for (top) the $d$ = 420 nm and (bottom) the $d$ = 380 nm array. In both color maps a strong connection between the time and temperature is observed as stripe like features. This implies that a given magnetization value can be found using several combinations of time and temperature. 
	The contour density along the gray lines for $t$ = 300 s corresponds to the relaxation rates shown in Fig. \ref{fig:Relax}d. The maps futher highlight the effect of the observation time on the temperature shift of the maximum for the relaxation rate, as indicated by the crosses for the case of $t$ = 3 s and $t$ = 300 s.}
	\label{fig:RelaxCont}
\end{figure}

The behavior in Fig. \ref{fig:Relax}a and b illustrates the interdependence of the remanent magnetization on time and temperature. This is further elucidated in Fig. \ref{fig:RelaxCont} were the magnetization is plotted as a function of temperature and time in a color map. It can be seen that the magnetization changes in a stripe like fashion, where a given value of the magnetization is not unique, but can be achieved through different combinations of time and temperature. The contour lines in the figure have constant magnetization and are separated by the same magnetization step. This implies that the time and temperature dependence of the relaxation rate $S$ is reflected in the density of contour lines, where a low density means a low relaxation rate. An alternative way of looking at the data presented in Fig. \ref{fig:Relax}d is to look at contour line density along the gray lines at $t$=300 s in Fig. \ref{fig:RelaxCont}. If the observation time is changed from 300s to a shorter observation time of 3 s, represented by the gray lines at $t$=3~s, the temperature of the maximum relaxation rate, increases by roughly 20 K. This accords with the shift of the M vs. T curves for different observation times in Fig. \ref{fig:AFM_MvsTMPMS}c and Fig. \ref{fig:Relax}c.

\begin{figure}[h!]
	\centering
		\includegraphics[width=0.45\textwidth]{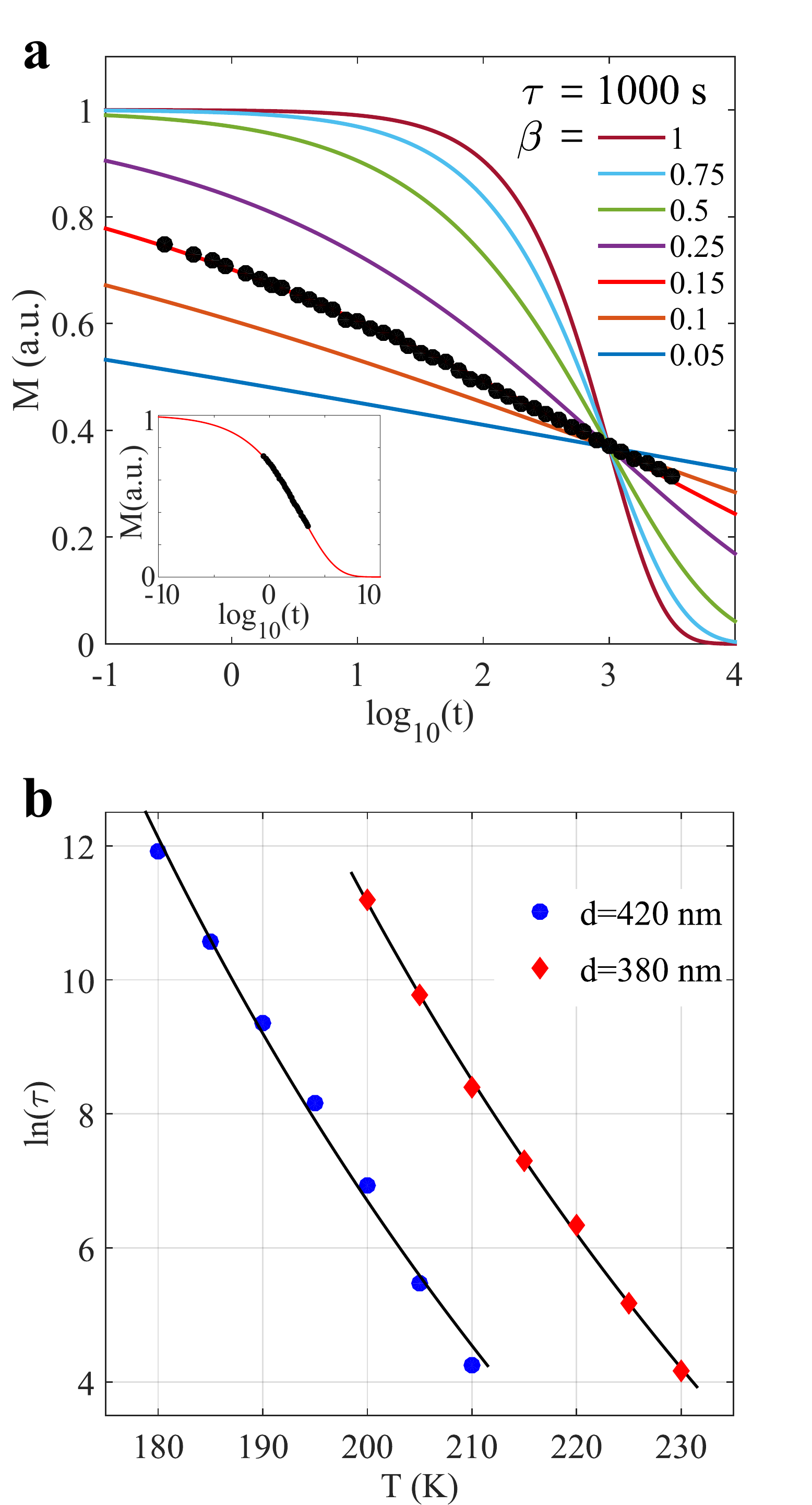}
	\caption{ \textbf{Stretched exponentials and fitting of \boldmath{$M_{TRM}(t)$ and $\tau(T)$}}. \textbf{a}, The variation of a stretched exponential, $M=M_0 exp[-(t/\tau)^\beta$], for different values of $\beta$ using a relaxation time  $\tau=1000$ s. The black dots correspond to the recorded relaxation data for the $d$~=~420 nm array at 200 K. In the inset the same data is fitted to a stretched exponential with $\beta \approx 0.15$ and $\tau \approx 1000$ s, in the time window $10^{-10}$ (0.1 ns) to  $10^{10}$ s ($\approx$ 300 years) which covers the majority of the relaxation from $M_0$ to zero magnetization. The magnetization is normalized to $M_0$, which is the magnetization directly after the field is switched off ($t~=~t_0$), corresponding to a fully dressed state of the array, see Fig. \ref{fig:Intro} (I). \textbf{b}, A fit of the temperature dependence of the relaxation time $\tau$ to the Vogel Fulcher law, $\tau=\tau_0 exp(E_B/k_B(T-T_0))$, for both arrays, yielding an average energy barrier of $E_B/k_B$ = 4500 K.
}
	\label{fig:Stretched_exponential}
\end{figure}

The relaxation in Fig. \ref{fig:Relax} occurs across a large time window and it is not described by a simple exponential function. However, a stretched exponential, $M=M_0$ $exp[-(t/\tau)^\beta$], where $\beta$ is the stretching exponent, $M_0$ is the magnetization at $t$ = 0 and $\tau$ is the relaxation time, can be used to describe the relaxation process \cite{Revell_NatPhys_2012,June2013}. To illustrate the influence of $\beta$, the stretched exponential decay in the time window 0.1 to 10 000 s is plotted in Fig.~\ref{fig:Stretched_exponential}a, using $\tau$ = 1000 s and $M_0$ = 1. As can be seen in the figure, a simple exponential ($\beta$=1) decays rapidly and does not describe the relaxation behavior illustrated in Fig. \ref{fig:Relax}. In Fig.~\ref{fig:Stretched_exponential}a the relaxation data measured at 200 K is shown for the 420 nm array;  comparing this to the stretched exponential curves one observes that it accurately matches the $\beta$~=~0.15 curve. The 200 K data was also fitted to a stretched exponential decay, yielding $\tau$~$\approx$~1000 s and $\beta$~$\approx$~0.15; red line in the inset of Fig. \ref{fig:Stretched_exponential}a. 
 Extending this fitting procedure to the measured relaxation data, presented in Fig. \ref{fig:Relax}a and b, the temperature evolution of $\tau$ can be studied. It should be noted that $\tau$ and $\beta$ are strongly coupled and therefore physical meaningful fits can only be made in a relatively narrow temperature region (see supplementary material for more details). For a region of about 30 K, $\beta$ is relative stable (around 0.15 for both arrays) and a reliable determination of the temperature dependence of $\tau$ can be made.
For example, $\tau$ changes by 3 orders of magnitude from about $10^5$ to $10^2$ s for both arrays, in the interval 180 to 210 K for the $d$ = 420 nm array and 200 to 230 K for the $d$~=~380 nm array.
 The extracted $\tau (T)$ values were fitted to the Vogel-Fulcher law \cite{Shtrikman1981,Calero-DiazDelCastillo2010}, $\tau=\tau_0exp[E_B/k_B (T-T_0)]$, where $\tau_0=10^{-11}$ s is a constant that describes the relaxation time of the individual islands at very high temperatures, $T_0$ describes the interaction strength, $E_B$ is the energy barrier which the magnetic moment of the island has to overcome in order to reverse and $k_B$ is the Boltzmann constant.  The fits for both arrays are shown in Fig. \ref{fig:Stretched_exponential}b. From these fits, the energy barrier, $E_B$, and the interaction strength $T_0$, are derived.  A value of $E_B/k_B$ = 4650 K is received for the 420 nm array and 4430 K for the 380 nm array, while for $T_0$ the values are 72 and 62 K, respectively. The results for $T_0$ are consistent with a stronger interaction between the elements in the 380 nm array as compared to the elements in the 420 nm array, as expected. 
The energy barrier of the islands is expected to be rather similar for the two arrays, due to the fact that the islands have the same size and shape in the two arrays and the temperature ranges for the fits partly overlap. Therefore fits where the energy barrier was fixed to 4500 K were also made, yielding $T_0$=77 K for the 380 nm array and $T_0$ = 60 K for the 420 nm array. The individual blocking temperature, $T_B$, of an island can be determined from $\tau_{obs}=\tau_0$ $exp[E_B/k_B T_B]$, where $\tau_{obs}$ is the observation time and is taken as 30 s. Using $E_B/k_B = 4500$ K a blocking temperature of approximately 160 K can be estimated. The energy barrier, $E_B$, of the individual islands can mainly be attributed to their shape anisotropy. Using this approximation the energy barrier can be estimated from $E_B(T) = \mu_o M_s(T)^2 \Delta N V/2$, where $\mu_o$ is the vacuum magnetic permeability, $M_s(T)$ is the magnetization of an island at temperature $T$, $\Delta N$ is the differential demagnetizing factor calculated using the Osborn methodology \cite{Osborn1945PhysRev} (see supplementary materials for details), and $V$ is the volume of the magnetic island. By using this approach and the temperature dependence of M described in the Methods, a mean value of the energy barrier was determined to be $E_B /k_B = 4900$ K at 195 K and $E_B /k_B = 4500$ K at 215 K. By comparing these calculated values of $E_B$ with the values estimated from the relaxation experiments, it can be seen that they are in good agreement. 

In summary, it is shown that the relaxation behavior of the investigated square artificial spin ice systems is well described by a stretched exponential decay. The arrays offer direct control over the relevant interaction - the magnetic dipolar interaction - via the geometry of the islands and the periodicity, while avoiding complicating factors such as disorder and magnetic domain walls. The relaxation experiments performed in this study yield direct information about the collective dynamics of the system over a quite wide time window and large temperature range. With the aid of data treatment based on well-established relaxation models, information about the energetics and dynamics of the individual building blocks can be acquired. Consequently, artificial spin ice arrays can be used as a new type of model system for the study of relaxation phenomena in magnetic nanosystems.

Concerning the collective dynamics of mesoscale magnetic systems, much remains to be explored, especially with respect to the temporal dynamics and phase transitions\cite{Anghinolfi_2015_NatComm}. Artificial spin ice and more general artificial ferroic systems are micro-magnetic variants of interacting many-body systems\cite{Heyderman_2013_review}. These have developed over the past years from a testbed for the study of ground-state ordering and low-energy behavior, to a playground for the exploration of collective excitations and dynamics\cite{Kapaklis2014, Gilbert2016_NatPhys}. Artificial ferroic systems with tunable dynamical properties can be designed by considering different materials and geometries rather than being limited just by microscopic material properties.
 This facilitates the study of dynamics in many-body systems on adjustable length, energy and time scales, opening up possibilities for the design of new devices with tailored electromagnetic properties in the relevant frequency regimes\cite{Gliga_2013_PRL}.

\section{\label{Method}Methods}

The magnetic arrays investigated here were produced by post-patterning on a $\delta$-doped Pd(Fe) thin film using e-beam lithography\cite{Kapaklis2014}. $\delta$-doped Pd(Fe) is a three-layer system of Pd (40 nm) - Fe (2.2 monolayers)- Pd (2 nm) produced by DC magnetron sputtering on a Magnesium oxide (MgO) substrate with a Vanadium (V) seeding-layer\cite{Parnaste2007, Kapaklis2014}. The two square artificial spin ice arrays used cover an area of 2 $\times$ 2 mm$^2$. The elongated islands forming the arrays are stadium shaped and have dimensions of $330~\rm{nm}$ $\times$ $150~\rm{nm}$. However the periodicity, $d$, is not the same, with one array having $d$~=~380 nm and the other $d$~=~420 nm, see Fig. \ref{fig:AFM_MvsTMPMS}a.

Magnetic characterization was performed using a custom built SQUID suitable for low field and low signal measurements \cite{Magnusson1997} and a commercial Quantum design MPMS SQUID magnetometer. In both set-ups the arrays were oriented in the [110]-direction with respect to the applied magnetic field, as shown in Fig. \ref{fig:AFM_MvsTMPMS}a. Magnetization as a function of temperature experiments where made using the temperature dependent thermoremanent magnetization protocol, $M_{TRM}(T)$, where the array is cooled from the superparamagnetic regime (300 K) in an applied magnetic field, $H_a$, down to a low temperature where the field is switched off and the magnetization is recorded upon heating with a constant rate. In the custom built SQUID an applied field of $H_a$=800 A/m (10 Oe) was used, while in the MPMS applied fields of $H_a$=800, 1600, 4000 and 8000 A/m (10, 20, 50 and 100 Oe) were used. Magnetization as a function of time was measured using the time dependent thermoremanent magnetization relaxation protocol, $M_{TRM}(t)$, where the array is cooled from the superparamagnetic regime in an applied field of 800 A/m (10 Oe) to the measurement temperature, $T_{m}$. Upon reaching $T_{m}$  the field is switched off and the magnetization is recorded as a function of time, $t$. $t=0$ is defined as the instant when the field is switched off. The experiment was made for several temperatures [$T_{m}$ = 170 to 260 K] and the relaxation was measured in the time window 0.3 to 3000 s. The magnetization in Fig. \ref{fig:Relax} and Fig. \ref{fig:RelaxCont} is normalized to $M_{TRM}$(T=130 K) for the corresponding array and is denoted as $M_{norm}$.

The magnetic moment of an individual island was determined from its area and the magnetization of the continuous film, measured by using the commercial MPMS SQUID magnetometer. The resulting island magnetic moment, determined in this way, is $m_0 = 7.4 \times 10^6$ $\mu_B$ at $T$ = 5 K. From measurements performed with a Vibrating Sample Magnetometer (VSM), it was revealed that the temperature dependence of the in-field volume magnetization can be described by the function $M_s(T) = m_0/V\cdot(1-T/T^*)^{0.5}$, with $T^*$ = 410 K and $V$ being the island volume. For the energy barrier calculation, $E_B$, a reduced island magnetic moment was used, in order to take into account the non-collinearities of the magnetic moments close to the edges of the islands.

\section{\label{Author contributions}Author contributions}

V.K. and B.H. initiated the work. V.K., M.S.A., E.\"O and R.M. planned the experiments. V.K., E.\"O. and M.S.A. carried out preliminary measurements, H.S., S.P. and A.S. fabricated the samples. M.S.A. performed the relaxation experiments. M.S.A. analyzed the data with contributions from P.N., R.M., H.S., S.P., E.\"O, B.H. and V.K.. M.S.A., B.H. and V.K. co-wrote the paper. All authors contributed to discussing the results and writing the paper.

\section{\label{Acknowledgements}Acknowledgements}
The authors thank Dr. Volker Neu for VSM measurements on the continuous $\delta$-doped Pd(Fe) thin films, Dr. Rimantas Brucas for sample treatment and Dr. Petra E. J\"onsson for discussions and advice concerning the preliminary SQUID measurements. The authors acknowledge support from the Knut and Alice Wallenberg Foundation, the Swedish Research Council and the Swedish Foundation for International Cooperation in Research. The patterning was performed at the Center for Functional Nanomaterials (CFN), Brookhaven National Laboratory, which is supported by the U.S. Department of Energy, Office of Basic Energy Sciences, under Contract No. DE-AC02-98CH10886.

\bibliographystyle{natphys}

\begin{thebibliography}{10}

\bibitem{Wang:2006kta}
R.~F. Wang, C.~Nisoli, R.~S. Freitas, J.~Li, W.~Mcconville, B.~J. Cooley, M.~S.
  Lund, N.~Samarth, C.~Leighton, V.~H. Crespi, and P.~Schiffer, {Artificial
  `spin ice' in a geometrically frustrated lattice of nanoscale ferromagnetic
  islands}, \textit{Nature} \textbf{439}, 303 (2006).

\bibitem{Kohlrausch}
R.~Kohlrausch, Theorie des elektrischen R\"uckstandes in der Leidener Flasche,
  \textit{Ann. Phys. (Leipzig)} \textbf{12}, 393 (1847).

\bibitem{Vogel}
H.~Vogel, The law of relation between viscosity of liquids and the temperature,
  \textit{Phys. Z.} \textbf{22}, 645 (1921).

\bibitem{Fulcher}
G.~S. Fulcher, Analysis of recent measurements of the viscosity of glasses,
  \textit{J. Am. Ceram. Soc.} \textbf{8}, 339 (1925).

\bibitem{Williams:1970ki}
G.~Williams and D.~C. Watts, {Non-symmetrical dielectric relaxation behaviour
  arising from a simple empirical decay function}, \textit{Transactions of the
  Faraday Society} \textbf{66}, 80--85 (1970).

\bibitem{Nature_dielectric_relaxation_1977}
A.~K. Jonscher, {The universal dielectric response}, \textit{Nature}
  \textbf{267}, 673--679 (1977).

\bibitem{Anderson_PRL_1984}
R.~G. Palmer, D.~L. Stein, E.~Abrahams, and P.~W. Anderson, {Models of
  Hierarchically Constrained Dynamics for Glassy Relaxation}, \textit{Physical
  Review Letters} \textbf{53}, 958--961 (1984).

\bibitem{Bohmer_1993}
R.~B{\"o}hmer, K.~L. Ngai, C.~A. Angell, and D.~J. Plazek, {Nonexponential
  relaxations in strong and fragile glass formers}, \textit{The Journal of
  Chemical Physics} \textbf{99}, 4201--4209 (1993).

\bibitem{Schoenholz:2016fb}
S.~S. Schoenholz, E.~D. Cubuk, D.~M. Sussman, E.~Kaxiras, and A.~J. Liu, A
  structural approach to relaxation in glassy liquids, \textit{Nature Physics}
  \textbf{12}, 469--471 (2016).

\bibitem{Weeks:2000dw}
E.~R. Weeks, J.~C. Crocker, A.~C. Levitt, A.~Schofield, and D.~A. Weitz,
  {Three-Dimensional Direct Imaging of Structural Relaxation Near the Colloidal
  Glass Transition}, \textit{Science} \textbf{287}, 627--631 (2000).

\bibitem{Andersson2014EPL}
M.~S. Andersson, J.~A.~D. Toro, S.~S. Lee, R.~Mathieu, and P.~Nordblad, Ageing
  dynamics of a superspin glass, \textit{EPL (Europhysics Letters)}
  \textbf{108}, 17004 (2014).

\bibitem{PARAK1991362}
F.~Parak and G.~U. Nienhaus, Glass-like behaviour of proteins as seen by
  \relax{M}\"ossbauer spectroscopy, \textit{Journal of Non-Crystalline Solids}
  \textbf{131}, 362 -- 368 (1991).

\bibitem{Ngai}
K.~L. Ngai, Relaxation and Diffusion in Complex Systems (Springer-Verlag New
  York, 2011).

\bibitem{Bramwell2001Science}
S.~T. Bramwell and M.~J.~P. Gingras, {Spin Ice State in Frustrated Magnetic
  Pyrochlore Materials}, \textit{Science} \textbf{294}, 1495--1501 (2001).

\bibitem{Giblin_NatPhys_2011}
S.~R. Giblin, S.~T. Bramwell, P.~C.~W. Holdsworth, D.~Prabhakaran, and
  I.~Terry, {Creation and measurement of long-lived magnetic monopole currents
  in spin ice}, \textit{Nature Physics} \textbf{7}, 252 (2011).

\bibitem{Revell_NatPhys_2012}
H.~M. Revell, L.~R. Yaraskavitch, J.~D. Mason, K.~A. Ross, H.~M.~L. Noad, H.~A.
  Dabkowska, B.~D. Gaulin, P.~Henelius, and J.~B. Kycia, {Evidence of impurity
  and boundary effects on magnetic monopole dynamics in spin ice},
  \textit{Nature Physics} \textbf{9}, 34--37 (2012).

\bibitem{Heyderman_2013_review}
L.~J. Heyderman and R.~L. Stamps, {Artificial ferroic systems: novel
  functionality from structure, interactions and dynamics}, \textit{Journal of
  Physics: Condensed Matter} \textbf{25}, 363201 (2013).

\bibitem{Farhan_PRL_2013}
A.~Farhan, P.~M. Derlet, A.~Kleibert, A.~Balan, R.~V. Chopdekar, M.~Wyss,
  J.~Perron, A.~Scholl, F.~Nolting, and L.~J. Heyderman, {Direct Observation of
  Thermal Relaxation in Artificial Spin Ice}, \textit{Physical Review Letters}
  \textbf{111}, 057204 (2013).

\bibitem{Kapaklis2014}
V.~Kapaklis, U.~B. Arnalds, A.~Farhan, R.~V. Chopdekar, A.~Balan, A.~Scholl,
  L.~J. Heyderman, and B.~Hj{\"{o}}rvarsson, {Thermal fluctuations in
  artificial spin ice.}, \textit{Nat. Nanotechnol.} \textbf{9}, 514--9 (2014).

\bibitem{June2013}
R.~K. June, J.~P. Cunningham, and D.~P. Fyhrie, {A Novel Method for
  Curvefitting the Stretched Exponential Function to Experimental Data.},
  \textit{Biomed. Eng. Res.} \textbf{2}, 153--158 (2013).

\bibitem{Shtrikman1981}
S.~Shtrikman and E.~P. Wohlfarth, The theory of the Vogel-Fulcher law of spin
  glasses, \textit{Phys. Lett. A} \textbf{85}, 467--470 (1981).

\bibitem{Calero-DiazDelCastillo2010}
V.~L. {Calero-Diaz Del Castillo} and C.~Rinaldi, {Effect of sample
  concentration on the determination of the anisotropy constant of magnetic
  nanoparticles}, \textit{IEEE Trans. Magn.} \textbf{46}, 852--859 (2010).

\bibitem{Osborn1945PhysRev}
J.~A. Osborn, Demagnetizing Factors of the General Ellipsoid, \textit{Phys.
  Rev.} \textbf{67}, 351--357 (1945).

\bibitem{Anghinolfi_2015_NatComm}
L.~Anghinolfi, H.~Luetkens, J.~Perron, M.~G. Flokstra, O.~Sendetskyi, A.~Suter,
  T.~Prokscha, P.~M. Derlet, S.~L. Lee, and L.~J. Heyderman, {Thermodynamic
  phase transitions in a frustrated magnetic metamaterial}, \textit{Nature
  Communications} \textbf{6}, 8278 (2015).

\bibitem{Gilbert2016_NatPhys}
I.~Gilbert, Y.~Lao, I.~Carrasquillo, L.~O{\textquoteright}Brien, J.~D. Watts,
  M.~Manno, C.~Leighton, A.~Scholl, C.~Nisoli, and P.~Schiffer, {Emergent
  reduced dimensionality by vertex frustration in artificial spin ice},
  \textit{Nature Physics} \textbf{12}, 162--165 (2016).

\bibitem{Gliga_2013_PRL}
S.~Gliga, A.~K{\'a}kay, R.~Hertel, and O.~G. Heinonen, {Spectral Analysis of
  Topological Defects in an Artificial Spin-Ice Lattice}, \textit{Physical
  Review Letters} \textbf{110}, 117205 (2013).

\bibitem{Parnaste2007}
M.~P{\"{a}}rnaste, M.~Marcellini, E.~Holmstr{\"{o}}m, N.~Bock, J.~Fransson,
  O.~Eriksson, and B.~Hj{\"{o}}rvarsson, {Dimensionality crossover in the
  induced magnetization of Pd layers}, \textit{J. Phys. Condens. Matter}
  \textbf{19}, 246213 (2007).

\bibitem{Magnusson1997}
J.~Magnusson, C.~Djurberg, P.~Granberg, and P.~Nordblad, {A low field
  superconducting quantum interference device magnetometer for dynamic
  measurements}, \textit{Review of Scientific Instruments} \textbf{68}, 3761
  (1997).

\bibitem{Vansteenkiste2014AIPAdv}
A.~Vansteenkiste, J.~Leliaert, M.~Dvornik, M.~Helsen, F.~Garcia-Sanchez, and
  B.~{Van Waeyenberge}, {The design and verification of MuMax3}, \textit{AIP
  Adv.} \textbf{4}, 107133 (2014).

\end{thebibliography}

\end{document}